\documentclass[12pt]{article}
\newcommand{\al}{\alpha}
\newcommand{\ba}{\begin{array}}
\newcommand{\be}{\begin{equation}}
\newcommand{\bea}{\begin{eqnarray}}
\newcommand{\cc}{{\cal C}}
\newcommand{\ck}{{\cal K}}
\newcommand{\cj}{{\cal J}}
\newcommand{\cl}{{\cal L}}
\newcommand{\da}{\dagger}
\newcommand{\ddz}{\frac{d}{dz}}
\newcommand{\ddzt}{\frac{d^2}{dz^2}}
\newcommand{\ddzth}{\frac{d^3}{dz^3}}  
\newcommand{\ea}{\end{array}}
\newcommand{\ee}{\end{equation}}
\newcommand{\eea}{\end{eqnarray}}
\newcommand{\fr}{\frac}

\newcommand{\lb}{\label}
\newcommand{\ld}{\ldots}
\newcommand{\lcb}{\left\{}

\newcommand{\lmd}{\left|}
\newcommand{\lra}{\longrightarrow} 
\newcommand{\lrb}{\left(}
\newcommand{\lsb}{\left[}
\newcommand{\nn}{\nonumber}

\newcommand{\rmd}{\right|}
\newcommand{\rra}{\right\rangle}
\newcommand{\rcb}{\right\}}
\newcommand{\rrb}{\right)}
\newcommand{\rsb}{\right]}

\newcommand{\half}{\fr{1}{2}} 
\begin{document}

\begin{flushright}
IMSc/2001/49 \\ 
30 August 2001 
\end{flushright}

\baselineskip16pt

\begin{center}
{\Large \bf Three dimensional quadratic algebras\,: 
Some realizations and representations} 
\end{center}

\smallskip

\baselineskip14pt

\begin{center}
V. Sunil Kumar$^1$, B. A. Bambah$^2$\footnote{E-mail:~{\tt bbsp@uohyd.ernet.in}}, 
and R. Jagannathan$^3$\footnote{E-mail:~{\tt jagan@imsc.ernet.in}} \\ 
{\em $^{1,2}$School of Physics, University of Hyderabad \\ 
Hyderabad - 500046, India \\
$^3$The Institute of Mathematical Sciences \\ 
C.I.T. Campus, Tharamani, Chennai - 600113, India}  
\end{center}

\vspace{0.5cm}

\noindent
Four classes of three dimensional quadratic algebras of the type 
$\lsb Q_0 , Q_\pm \rsb$ $=$ $\pm Q_\pm$, 
$\lsb Q_+ , Q_- \rsb$ $=$ $aQ_0^2 + bQ_0 + c$, where $(a,b,c)$ are constants 
or central elements of the algebra, are constructed using a generalization 
of the well known two-mode bosonic realizations of $su(2)$ and $su(1,1)$.  
The resulting matrix representations and single variable differential operator 
realizations are obtained.  Some remarks on the mathematical and physical 
relevance of such algebras are given. 

\bigskip 

\section{Introduction} 
\renewcommand{\theequation}{\arabic{section}.{\arabic{equation}}}
\setcounter{equation}{0}
In recent times there has been a great deal of interest in non-linear  
deformations of Lie algebras because of their significant applications 
in several branches of physics.  This is largely based on the realization 
that the physical operators relevant for defining the dynamical algebra 
of a system need not be closed under a linear (Lie) algebra, but might 
obey a nonlinear algebra.  Such nonlinear algebras are, in general, 
characterized by commutation relations of the form 
\be
\lsb T_i , T_j \rsb = C_{ij} \lrb\lcb T_k \rcb\rrb\,, 
\lb{cr} 
\ee 
where the functions $\lcb C_{ij} \rcb$ of the generators $\lcb T_k \rcb$ are 
constrained by the Jacobi identity 
\be
\lsb T_i , C_{jk} \rsb + 
\lsb T_j , C_{ki} \rsb + 
\lsb T_k , C_{ij} \rsb = 0\,.  
\ee
The functions $\lcb C_{ij} \rcb$ can be infinite power series in 
$\lcb T_k \rcb$ as is in the case of quantum algebras (with further Hopf 
algebraic restrictions) and $q$-oscillator algebras.  When $\lcb C_{ij} \rcb$ 
are polynomials of the generators one gets the socalled polynomially nonlinear, 
or simply polynomial, algebras.  A special case of interest is when the 
commutation relations (\ref{cr}) take the form 
\be
\lsb T_i , T_j \rsb = c_{ij}^k T_k\,, \quad 
\lsb T_i , T_\al \rsb = t_{i\al}^\beta T_\beta\,, \quad 
\lsb T_\al , T_\beta \rsb = f_{\al\beta} \lrb\lcb T_k \rcb\rrb\,,  
\ee 
containing a linear subalgebra.  Simplest examples of such algebras occur 
when one gets a three dimensional algebra 
\be
\lsb N_0 , N_\pm \rsb = \pm N_\pm\,, \qquad
\lsb N_+ , N_- \rsb = f\lrb N_0 \rrb\,, 
\lb{fn} 
\ee 
in which $f\lrb N_0 \rrb$ is a polynomial in $N_0$.  In general, 
\be 
\cc = N_+ N_- + g \lrb N_0 - 1 \rrb = N_- N_+ + g \lrb N_0 \rrb  
\lb{genC}
\ee
is a Casimir operator of the algebra (\ref{fn}) where $g \lrb N_0 \rrb$ can be 
determined from the relation 
\be
g \lrb N_0 \rrb - g \lrb N_0 - 1 \rrb = f \lrb N_0 \rrb\,.
\lb{ggf}
\ee  

If $f\lrb N_0 \rrb$ is quadratic in $N_0$ we have a quadratic algebra and  
if $f\lrb N_0 \rrb$ is cubic in $N_0$ we have a cubic algebra.  The polynomial 
algebras, in particular the quadratic and cubic algebras, and their 
representations are involved in the studies of several problems in quantum 
mechanics, statistical physics, field theory, Yang-Mills type gauge theories, 
two dimensional integrable systems, etc. (\cite{L}-\cite{F}).  These algebras 
have also been found to occur in quantum optics with the observation that 
quantum optical Hamiltonians describing multiphoton processes have dynamical 
algebras described by polynomially deformed $su(2)$ and $su(1,1)$ algebras 
\cite{K}.  Coherent states of different kinds of nonlinear oscillator algebras 
have been presented by several authors (\cite{M}-\cite{Q2}).  Recently, a 
general unified approach for finding the coherent states of three dimensional  
polynomial algebras, has been presented by us \cite{Su}.  The particular examples 
of three dimensional quadratic algebras considered by us in \cite{Su} correspond 
to two of the four classes of algebras to be presented below.  Algebras of the 
type (\ref{fn}) have also been studied from a purely mathematical point of view 
\cite{Sm} suggesting a rich theory of representations for them.  Algebras of the 
type (\ref{fn}) with commutator $\lsb N_+ , N_- \rsb$ replaced by the 
anticommutator $\lcb N_+ , N_- \rcb$ leading to polynomial deformations of the 
superalgebra $osp(1|2)$ have also been investigated \cite{Jv}.  

As is well known, in the case of classical Lie algebras bosonic realizations 
play a very useful role in the representation theory and applications to 
physical problems.  The main purpose of this work is to study some aspects of 
three dimensional quadratic algebras relating to bosonic realizations and 
the associated matrix representations, differential operator realizations, 
and physical relevance.  In Section 2 we briefly review the well known 
construction of $su(2)$ and $su(1,1)$ algebras in terms of two-mode bosonic 
operators.  In Section 3 we show that a generalization of the Jordan-Schwinger 
method, using $su(2)$ or $su(1,1)$ and a boson algebra as the building blocks, 
leads to the construction of four classes of three dimensional quadratic 
algebras.  In Sections 4-7 we exhibit the three-mode bosonic realizations of 
these quadratic algebras and derive the associated matrix representations and 
single variable differential operator realizations. Finally, in Section 8 we 
conclude with a few remarks on the physical and mathematical relevance of these 
quadratic algebras.  
 
\section{Two-mode bosonic construction of $su(2)$ and $su(1,1)$\,: A brief 
review} 
\renewcommand{\theequation}{\arabic{section}.{\arabic{equation}}}
\setcounter{equation}{0} 
Let us briefly recall the study of $su(2)$ and $su(1,1)$ in terms of two-mode 
bosonic realizations, to fix the framework and notations for our work.  Let 
$\lrb a_1 , a_1^\da \rrb$ and $\lrb a_2 , a_2^\da \rrb$ be two mutually 
commuting boson annihilation-creation operator pairs.  Let 
$H_1 = N_1+\half = a_1^\da a_1+\half$ and 
$H_2 = N_2+\half = a_2^\da a_2+\half$.  As is well known, 
$\lrb J_0, J_+, J_- \rrb$ defined by 
\be
J_0 = \half \lrb H_1 - H_2 \rrb\,, \quad 
J_+ = a_1^\da a_2\,, \quad
J_- = J_+^\da = a_1 a_2^\da\,, 
\lb{js}
\ee 
satisfy the $su(2)$ algebra, 
\be
\lsb J_0 , J_\pm \rsb = \pm J_\pm\,, \quad
\lsb J_+ , J_- \rsb = 2J_0\,.  
\ee
In this Jordan-Schwinger realization of $su(2)$, $H_1 + H_2$ is seen to be a 
central element\,: if 
\be
\cl = \half \lrb H_1 + H_2 \rrb\,, 
\ee
then,
\be
\lsb \cl , J_{0,\pm} \rsb = 0\,.  
\ee
The usual Casimir operator is  
\be 
\cc = J^2 = J_+J_- + J_0 \lrb J_0 - 1 \rrb = \cl^2 - \fr{1}{4}\,. 
\ee
Consequently, the application of the realization (\ref{js}) on a set of 
$2j+1$ two-mode Fock states $\lmd n_1 \rra \lmd n_2 \rra$, with constant 
$n_1+n_2$ $=$ $2j$, leads to the $(2j+1)$-dimensional unitary irreducible 
representation for each $j = 0,1/2,1,\ld\,$. Thus, with  
$\lcb \lmd j,m \rra \right.$ $=$ $\lmd j+m \rra\lmd j-m \rra$ 
$\lmd m = j,j-1,\ld\,,-j \rcb$ as the basis states, one gets the $j$-th unitary 
irreducible representation 
\bea
J_0 \lmd j , m \rra & = & m \lmd j , m \rra\,, \nn \\ 
J_\pm \lmd j , m \rra & = & 
      \sqrt{(j \mp m)(j \pm m+1)}\,\lmd j , m \pm 1 \rra\,, \nn \\ 
\cl\lmd j , m \rra & = & \lrb j+\half \rrb \lmd j , m \rra\,, \quad 
J^2\lmd j , m \rra = j(j+1)\lmd j , m \rra\,, \nn \\  
   &  & \qquad \qquad \qquad \qquad m = j,j-1,\ld\,,-j\,.  
\lb{su2rep}
\eea 

Let us now consider the single variable differential operator realization 
corresponding to the above matrix representation (\ref{su2rep}).  With the 
Fock-Bargmann correspondence 
\be
\lrb a^\da , a \rrb \lra \lrb z , \ddz \rrb\,, \qquad
\lmd n \rra \lra \fr{z^n}{\sqrt{n!}}\,,
\ee
we can make the association 
\bea
\lmd j , m \rra & \lra & \fr{z_1^{j+m} z_2^{j-m}}{\sqrt{(j+m)!(j-m)!}} 
  = \fr{z_2^{2j} \lrb z_1/z_2 \rrb^{j+m}}{\sqrt{(j+m)!(j-m)!}}\,, \nn \\
  &  &  \qquad \qquad m = -j, -j+1, \ld\,, j-1, j\,.
\eea  
Since $j$ is a constant for a given representation we can rewrite the above 
as a mapping to monomials 
\be 
\lmd j , m \rra \lra \psi_{j,n}(z) = \fr{z^n}{\sqrt{n!(2j-n)!}}\,, \quad 
n = 0, 1, 2, \ld\,, 2j\,. 
\lb{mono} 
\ee 
Then, it is obvious that the above set of $(2j+1)$ monomials (\ref{mono}) 
forms the basis carrying the finite dimensional representation (\ref{su2rep}) 
corresponding to the single variable realization 
\be 
J_0 = z \ddz - j\,, \quad 
J_+ = -z^2 \ddz + 2jz\,, \quad 
J_- = \ddz\,. 
\ee

In an analogous way, $\lrb K_0, K_+, K_- \rrb$ defined by  
\be
K_0 = \half \lrb H_1 + H_2 \rrb\,, \quad
K_+ = a_1^\da a_2^\da\,, \quad
K_- = K_+^\da = a_1 a_2\,,  
\lb{su11}
\ee
satisfy the $su(1,1)$ algebra  
\be
\lsb K_0 , K_\pm \rsb = \pm K_\pm\,, \qquad 
\lsb K_+ , K_- \rsb = -2K_0\,.
\ee 
Now, 
\be
\cl = \half \lrb H_1 - H_2 \rrb 
\ee
is a central element of the algebra\,:
\be
\lsb \cl , K_{0,\pm} \rsb = 0\,.  
\ee 
The usual Casimir operator is 
\be 
\cc = K^2 = K_+ K_- - K_0 \lrb K_0-1 \rrb = \fr{1}{4} - \cl^2\,. 
\ee 
Consequently, the application of the realization (\ref{su11}) on any infinite 
set of two-mode Fock states 
$\lcb \lmd k , n \rra = \lmd n+2k-1 \rra\lmd n \rra \right.$ 
$\lmd n = 0,1,2,\,\ld \rcb$, with constant $n_1 - n_2 = 2k-1$, leads to the 
infinite dimensional unitary irreducible representation, the so-called positive 
discrete representation ${\cal D}^+(k)$\,, corresponding to any  
$k = 1/2, 1, 3/2,\,\ld\,.$\,: 
\bea
K_0 \lmd k , n \rra & = & \lrb k+n \rrb \lmd k , n \rra\,, \nn \\
K_+ \lmd k , n \rra & = & \sqrt{(2k+n)(n+1)}\,\lmd k , n+1 \rra\,, \nn \\
K_- \lmd k , n \rra & = & \sqrt{(2k+n-1)n}\,\lmd k , n-1 \rra\,, \nn \\
\cl\lmd k , n \rra & = & \lrb k - \half \rrb \lmd k , n \rra\,, \quad 
K^2\lmd k , n \rra = k(1-k)\lmd k , n \rra\,, \nn \\ 
  &  & \qquad \qquad \qquad \qquad n = 0,1,2,\,\ld\,. 
\lb{su11rep}
\eea 
Note that the choice of basis states as 
$\lcb \lmd k , n \rra = \lmd n \rra\lmd n+2k-1 \rra \right.$ 
$\lmd n = 0,1,2,\,\ld \rcb$, with $n_1-n_2 = 1-2k$, is also posssible leading to 
the same representation (\ref{su11rep}), with $K^2 = k(1-k)$, but corresponding 
to $\cl = \half - k$.    

As in the $su(2)$ case, we can make the association 
\be
\lmd k , n \rra \lra \fr{z_1^{n+2k-1}z_2^n}{\sqrt{(n+2k-1)!n!}} 
     = \fr{\lrb z_1 z_2 \rrb^n z_1^{2k-1}}{\sqrt{(n+2k-1)!n!}}\,.
\ee
Then, with $k$ being constant in a given representation, it is obvious that 
the infinite set of monomials 
\be
\phi_{k,n}(z) = \fr{z^n}{\sqrt{(n+2k-1)!n!}}\,, \quad 
n = 0,1,2,\,\ld\,, 
\ee
forms the basis carrying the representation (\ref{su11rep}) corresponding to 
the single variable realization 
\be
K_0 = z \ddz + k\,, \quad 
K_+ = z\,, \quad 
K_- = z \ddzt + 2k \ddz\,.
\ee

\section{Construction of four classes of three dimensional quadratic algebras} 
\renewcommand{\theequation}{\arabic{section}.{\arabic{equation}}}
\setcounter{equation}{0}
A three dimensional quadratic algebra is defined, in general, by the 
commutation relations 
\be
\lsb Q_0 , Q_\pm \rsb = \pm Q_\pm\,, \quad 
\lsb Q_+ , Q_- \rsb = a Q_0^2 + b Q_0 + c\,, 
\lb{genalg}
\ee 
where the structure constants $(a,b,c)$ are constants, or central elements of 
the algebra so that they take constant values in any irreducible representation.  
In general, a Casimir operator of the algebra (\ref{genalg}) is 
\be
\cc = Q_+Q_- + \fr{1}{3}aQ_0^3 - \half(a-b)Q_0^2 
             + \fr{1}{6}(a-3b+6c)Q_0 - c\,.
\lb{casimir} 
\ee
When $a = c = 0$ and $b = \pm 2$ one gets, respectively, $su(2)$ and $su(1,1)$ 
as special cases.  

Quadratic algebras and their representations, physically relevant to the 
situations, have been studied in physics literature in the context of several 
specific physical problems where $(a, b, c)$ take particular values depending on 
the problem studied.  But, there seems to be no systematic study of realizations, 
representations, and any classification of the quadratic algebras corresponding 
to arbitrary values of $(a, b, c)$.  The main purpose of this work is to observe 
that a generalization of the Jordan-Schwinger method leads to four classes of 
three dimensional quadratic algebras in each of which the constants $(a, b, c)$ 
take particular series of values.  To this end we proceed as follows.  

Let $\lcb \left. \lrb a_i , a_i^\da \rrb \rmd i = 1,2,3 \rcb$ be three mutually 
commuting boson annihilation-creation operator pairs.  Let $N_i = a_i^\da a_i$ 
and $H_i = N_i + \half$ for each $i = 1,2,3$.  Now, we let 
\be
Q_0 = \half \lrb J_0 - N_3 \rrb\,, \qquad 
Q_+ = J_+ a_3\,, \qquad Q_- = J_- a_3^\da\,, 
\lb{q2-def}
\ee
replacing in (\ref{js}) $\lrb H_1, a_1^\da, a_1 \rrb$ by 
$\lrb J_0, J_+, J_- \rrb$ respectively, and $\lrb H_2, a_2^\da, a_2 \rrb$ by 
$\lrb N_3, a_3^\da, a_3 \rrb$ respectively (note that $\lrb J_0, J_+, J_- \rrb$ 
already contain two sets of bosonic operators).  It is now straightforward to 
observe that, with $\cl$ and $\cj$ defined by 
\be 
\cl = \half\lrb J_0 + N_3 \rrb\,, \qquad
\cj = J_+J_- + J_0\lrb J_0 - 1 \rrb = J^2\,, 
\ee
we have 
\bea 
\lsb Q_0 , Q_\pm \rsb & = & \pm Q_\pm\,, \nn \\
\lsb Q_+ , Q_- \rsb & = & -3Q_0^2 - \lrb 2\cl - 1\rrb Q_0 
                          + \lrb \cj + \cl \lrb\cl+1\rrb \rrb\,, 
\lb{q2-}
\eea  
and
\be
\lsb \cl , \cj \rsb = 0\,, \qquad 
\lsb \cl , Q_{0,\pm} \rsb = 0\,, \qquad 
\lsb \cj , Q_{0,\pm} \rsb = 0\,.
\ee 
The Casimir operator (\ref{casimir}) becomes  
\be
\cc = Q_+Q_- - Q_0^3 - (\cl-2)Q_0^2 + \lrb \cj+\cl^2+2\cl-1 \rrb Q_0 
             - \lrb \cj + \cl \lrb\cl+1\rrb \rrb\,. 
\lb{q2-co} 
\ee

Next, let us replace in (\ref{su11}) $\lrb H_1, a_1^\da, a_1 \rrb$ by 
$\lrb J_0, J_+, J_- \rrb$ respectively, and $\lrb H_2, a_2^\da, a_2 \rrb$ by 
$\lrb N_3, a_3^\da, a_3 \rrb$ respectively.  This leads to the definitions 
\be
Q_0 = \half \lrb J_0 + N_3 \rrb\,, \qquad 
Q_+ = J_+ a_3^\da\,, \qquad Q_- = J_- a_3\,,  
\lb{q2+def}
\ee
which obey the algebra 
\bea 
\lsb Q_0 , Q_\pm \rsb & = & \pm Q_\pm\,, \nn \\
\lsb Q_+ , Q_- \rsb & = & 3Q_0^2 + \lrb 2\cl + 1\rrb Q_0 
                          - \lrb \cj + \cl\lrb\cl-1\rrb\rrb\,,  
\lb{q2+}
\eea  
with 
\be
\cl = \half\lrb J_0 - N_3 \rrb\,, \qquad
\cj = J_+J_- + J_0\lrb J_0 - 1 \rrb = J^2  
\ee
being the central elements of the algebra.  The Casimir operator becomes  
\be
\cc = Q_+Q_- + Q_0^3 + (\cl-1)Q_0^2 - \lrb \cj+\cl^2 \rrb Q_0 
             + \lrb \cj + \cl\lrb\cl-1\rrb \rrb\,. 
\lb{q2+co} 
\ee

Let us now use the $su(1,1)$ generators instead of the $su(2)$ generators in 
the above scheme.  Thus we replace in (\ref{q2-def}) $(J_0, J_+, J_-)$ by 
$(K_0, K_+, K_-)$ respectively.  The result is that 
\be
Q_0 = \half \lrb K_0 - N_3 \rrb\,, \qquad 
Q_+ = K_+ a_3\,, \qquad Q_- = K_- a_3^\da\,, 
\ee
satisfy the algebra 
\bea 
\lsb Q_0 , Q_\pm \rsb & = & \pm Q_\pm\,, \nn \\ 
\lsb Q_+ , Q_- \rsb & = & 3Q_0^2 + (2\cl-1) Q_0 
                        + \lrb \ck - \cl\lrb\cl+1\rrb \rrb\,, 
\lb{q11-} 
\eea 
with 
\be
\cl = \half \lrb K_0 + N_3 \rrb\,, \qquad 
\ck = K_+ K_- - K_0 \lrb K_0-1 \rrb = K^2\,, 
\ee
as the central elements.  The Casimir operator becomes  
\be
\cc = Q_+Q_- + Q_0^3 + (\cl-2)Q_0^2 + \lrb \ck-\cl^2-2\cl+1 \rrb Q_0 
             - \lrb \ck - \cl\lrb\cl+1\rrb \rrb\,. 
\lb{q11-co} 
\ee 

If we replace in (\ref{q2+def}) $(J_0, J_+, J_-)$ by $(K_0, K_+, K_-)$
respectively, the result is that 
\bea
Q_0 = \half \lrb K_0 + N_3 \rrb\,, \qquad 
Q_+ = K_+ a_3^\da\,, \qquad Q_- = K_- a_3\,, 
\eea 
satisfy the algebra 
\bea 
\lsb Q_0 , Q_\pm \rsb & = & \pm Q_\pm\,, \nn \\ 
\lsb Q_+ , Q_- \rsb & = & -3Q_0^2 - (2\cl + 1)Q_0 
                          - \lrb\ck - \cl\lrb\cl-1\rrb \rrb\,,
\lb{q11+} 
\eea 
with 
\be
\cl = \half \lrb K_0 - N_3 \rrb\,, \qquad 
\ck = K_+ K_- - K_0 \lrb K_0-1 \rrb = K^2\,, 
\ee
being the central elements.  The Casimir operator is 
\be
\cc = Q_+Q_- - Q_0^3 - (\cl-1)Q_0^2 - \lrb \ck-\cl^2 \rrb Q_0 
             + \lrb \ck - \cl\lrb\cl-1\rrb \rrb\,. 
\lb{q11+co} 
\ee

Thus, by combining the generators of $su(2)$, or $su(1,1)$, and an oscillator 
algebra, analogous to the way two oscillator algebras are combined to get 
$su(2)$ or $su(1,1)$, we get four classes of three dimensional quadratic  
algebras of the type (\ref{genalg}) with $(a, b, c)$ as central elements.  
Let us call the algebras (\ref{q2-}), (\ref{q2+}), (\ref{q11-}) and (\ref{q11+}), 
respectively, $Q^-(2)$, $Q^+(2)$, $Q^-(1,1)$ and $Q^+(1,1)$ where the 
superscripts $\pm$ indicate whether the corresponding $Q_0$ is a sum or a 
difference (correspondingly the forms of $Q_\pm$ get fixed).  We shall study 
the realizations and representations of these quadratic algebras in the following sections.  

\section{Representations of $Q^-(2)$} 
\renewcommand{\theequation}{\arabic{section}.{\arabic{equation}}}
\setcounter{equation}{0}
For the algebra $Q^-(2)$ defined in (\ref{q2-}) $\cj$ and $\cl$ are to be 
constants in any irreducible representation.  This implies that we can take the 
basis states of the irreducible representations as 
\be
\lmd j,l,m \rra = \lmd j,m \rra \lmd 2l-m \rra\,, 
\lb{q2-base}
\ee
such that 
\be
\cj\lmd j,l,m \rra = j(j+1)\lmd j,l,m \rra\,, \qquad 
\cl\lmd j,l,m \rra = l\lmd j,l,m \rra\,,
\ee
where $\lcb\lmd j,m \rra\rcb$ are the $su(2)$ basis states (\ref{su2rep}) and 
$\lcb\lmd 2l-m \rra\rcb$ are the oscillator eigenstates such that 
$N_3\lmd 2l-m \rra$ $=$ $(2l-m)\lmd 2l-m \rra$.  This requires $2l-m$ to be a 
nonnegative integer for all values of $m$.  Thus there arise two cases to be 
considered separately.  In general, the Casimir operator (\ref{q2-co}) has 
the value $(l+1)\lsb j(j+1)-l(l-1) \rsb$ in both the cases.   \\ 

\noindent 
{\it Case}-I\,: $2l-j \geq 0$

\medskip 

In this case the set of basis states 
\be
\lmd j,l,m \rra = \lmd j,m \rra \lmd 2l-m \rra\,, \quad 
m = -j, -j+1,\,\ld\,j-1, j\,,  
\ee
carry the $(j,l)$-th, $(2j+1)$-dimensional, unitary irreducible representation  
\bea
Q_0\lmd j,l,m \rra & = & (m-l)\lmd j,l,m \rra\,, \nn \\
Q_+\lmd j,l,m \rra & = & \sqrt{(j-m)(j+m+1)(2l-m)}\lmd j,l,m+1 \rra\,, \nn \\
Q_-\lmd j,l,m \rra & = & \sqrt{(j+m)(j-m+1)(2l-m+1)}\lmd j,l,m-1 \rra\,, \nn \\ 
  &   & \qquad \qquad m = -j, -j+1,\,\ld\,j-1, j\,.  
\lb{q2-rep1}
\eea 
Let us look at the two dimensional representations.  These correspond to 
$j = 1/2$ and for each value of $l$ $=$ $1/4,3/4,5/4,\,\ld\,,$ there is a 
two dimensional representation given by 
\bea 
Q_0 & = & \lrb \ba{cc}
               -\half-l & 0   \\
                     0   & \half-l \ea \rrb\,, \nn \\  
Q_+ & = & \lrb \ba{cc} 
                   0 & 0 \\
           \sqrt{2l+\half} & 0 \ea \rrb\,, \qquad 
Q_- = \lrb \ba{cc}
          0 & \sqrt{2l+\half}  \\
          0 & 0          \ea \rrb\,, \nn \\ 
\cj & = & 3/4\,, \qquad \cl = l\,, \qquad \cc = (-4l^3+7l+3)/4\,, 
\eea
as can be verified directly. \\

\noindent 
{\it Case}-II\,: $2l-j < 0$

\medskip 

In this case the set of basis states 
\be
\lmd j,l,m \rra = \lmd j,m \rra \lmd 2l-m \rra\,, \quad 
m = -j, -j+1,\,\ld\,2l\,,  
\ee
carry the $(j,l)$-th unitary irreducible representation of dimension $j+2l+1$ 
given by  
\bea
Q_0\lmd j,l,m \rra & = & (m-l)\lmd j,l,m \rra\,, \nn \\
Q_+\lmd j,l,m \rra & = & \sqrt{(j-m)(j+m+1)(2l-m)}\lmd j,l,m+1 \rra\,, \nn \\
Q_-\lmd j,l,m \rra & = & \sqrt{(j+m)(j-m+1)(2l-m+1)}\lmd j,l,m-1 \rra\,, \nn \\
  &   & \qquad \qquad m = -j, -j+1,\,\ld\,2l\,.  
\lb{q2-rep2} 
\eea 
In this case for any $j > 1/2$ there is a two dimensional representation 
corresponding to $l = (1-j)/2$.  Explicitly, 
\bea 
Q_0 & = & \half \lrb \ba{cc}
                         -j-1 & 0   \\
                           0  & 1-j \ea \rrb\,, \nn \\  
Q_+ & = & \lrb \ba{cc} 
                   0 & 0 \\
           \sqrt{2j} & 0 \ea \rrb\,, \qquad 
Q_- = \lrb \ba{cc}
          0 & \sqrt{2j}  \\
          0 & 0          \ea \rrb\,, \nn \\ 
\cj & = & j(j+1)\,, \quad \cl = (1-j)/2\,, \quad 
\cc = (-3j^3+5j^2+11j+3)/8\,, \nn \\ 
    &   &  
\eea
as can be verified directly. \\
  
By using the two-mode bosonic realization of $su(2)$ we can write down the 
three-mode bosonic realization of $Q^-(2)$ as 
\be
Q_0 = \fr{1}{4}\lrb a_1^\da a_1-a_2^\da a_2-2a_3^\da a_3 \rrb\,, \quad 
Q_+ = a_1^\da a_2a_3\,, \quad Q_- = a_1a_2^\da a_3^\da\,.
\lb{q2-bose}
\ee
Correspondingly we can take the basis states (\ref{q2-base}) of the irreducible 
representations as the three-mode Fock states 
\be
\lmd j,l,m \rra = \lmd j+m \rra\lmd j-m \rra \lmd 2l-m \rra\,. 
\ee 
Then the action of $\lrb Q_0,Q_+,Q_- \rrb$ defined by (\ref{q2-bose}) on these basis states 
leads to the irreducible representations (\ref{q2-rep1}) and (\ref{q2-rep2}), respectively, 
in the two cases $2l-j$ $\geq$ $0$ and $2l-j$ $<$ $0$.  

Let us now make the association
\be
\lmd j,m,l \rra \lra \fr{z_2^{2j}z_3^{2l+j}\lrb z_1/z_2z_3\rrb^{j+m}}
                     {\sqrt{(j+m)!(j-m)!(2l-m)!}}\,.
\ee
Since $j$ and $l$ are constants, this shows that the set of functions
\be
\psi^-_{j,l,n}(z) = \fr{z^n}{\sqrt{n!(2j-n)!(2l+j-n)!}}\,, \quad 
 n = 0,1,2,\,\ld\,,2j,\ {\mbox{or}}\ 2l+j\,, 
\ee
forms the basis for the representation (\ref{q2-rep1}) or (\ref{q2-rep2}), 
respectively, corresponding to the single variable realization 
\bea
Q_0 & = & z\fr{d}{dz}-j-l\,, \nn \\ 
Q_+ & = & z^3\fr{d^2}{dz^2}-(2l+3j+1)z^2\fr{d}{dz}+2j(2l-j)z\,, \quad
Q_- = \fr{d}{dz}\,.
\eea  

\section{Representations of $Q^+(2)$} 
\renewcommand{\theequation}{\arabic{section}.{\arabic{equation}}}
\setcounter{equation}{0}
For the algebra $Q^+(2)$ defined in (\ref{q2+}) the constancy of $\cj$ and 
$\cl$ in any irreducible representation implies that the basis states can be 
taken as 
\be
\lmd j,l,m \rra = \lmd j,m \rra \lmd m-2l \rra\,, 
\lb{q2+base}
\ee
such that 
\be
\cj\lmd j,l,m \rra = j(j+1)\lmd j,l,m \rra\,, \qquad 
\cl\lmd j,l,m \rra = l\lmd j,l,m \rra\,,
\ee
where $\lcb\lmd j,m \rra\rcb$ are the $su(2)$ basis states and 
$\lcb\lmd m-2l \rra\rcb$ are the oscillator eigenstates such that 
$N_3\lmd m-2l \rra$ $=$ $(m-2l)\lmd m-2l \rra$.  The requirement that $m-2l$ 
has to be a nonnegative integer for any value of $m$ implies that $2l+j$ has to 
be always zero or a negative integer.  Thus the irreducible representations of 
this algebra are labeled by the pairs $\lcb(j,l)\rcb$ such that $2l+j$ is 
zero or a negative integer.  The corresponding $(j,l)$-th irreducible 
representation is always $(2j+1)$-dimensional and is given by    
\bea
Q_0\lmd j,l,m \rra & = & (m-l)\lmd j,l,m \rra\,, \nn \\
Q_+\lmd j,l,m \rra & = & \sqrt{(j-m)(j+m+1)(m-2l+1)}\lmd j,l,m+1 \rra\,, \nn \\
Q_-\lmd j,l,m \rra & = & \sqrt{(j+m)(j-m+1)(m-2l)}\lmd j,l,m-1 \rra\,, \nn \\
  &   & \qquad \qquad m = -j, -j+1,\,\ld\,j-1, j\,.  
\lb{q2+rep}
\eea 
The Casimir operator (\ref{q2+co}) takes the value 
$(1-l)\lsb j(j+1)-l(l+1) \rsb$ in this representation.  Note that the 
irreducible representations in this case are analogous to those of $su(2)$ 
except that there are infinitely many inequivalent irreducible representations 
of the same dimension corresponding to the infinity of the possible values of 
$l$ for each value of $j$.  

The two dimensional representations correspond to $j = 1/2$ and $l$ $=$ 
$-1/4,-3/4,-5/4,\,\ld\,,$ and are given by 
\bea 
Q_0 & = & \lrb \ba{cc}
                -\half-l & 0   \\
                          0  & \half-l \ea \rrb\,, \nn \\  
Q_+ & = & \lrb \ba{cc} 
                   0 & 0 \\
           \sqrt{\half-2l} & 0 \ea \rrb\,, \qquad 
Q_- = \lrb \ba{cc}
          0 & \sqrt{\half-2l}  \\
          0 & 0          \ea \rrb\,, \nn \\ 
\cj & = & 3/4\,, \qquad \cl = l\,, \qquad 
\cc = \fr{1}{4}\lrb 4l^3-7l+3 \rrb\,.  
\eea

By using the two-mode bosonic realization of $su(2)$ we can write down the 
three-mode bosonic realization of $Q^+(2)$ as 
\be
Q_0 = \fr{1}{4}\lrb a_1^\da a_1-a_2^\da a_2+2a_3^\da a_3 \rrb\,, \quad 
Q_+ = a_1^\da a_2a_3^\da\,, \quad Q_- = a_1a_2^\da a_3\,.
\lb{q2+bose}
\ee
Correspondingly we can take the basis states (\ref{q2+base}) of the irreducible 
representations as the three-mode Fock states 
\be
\lmd j,l,m \rra = \lmd j+m \rra\lmd j-m \rra \lmd m-2l \rra\,. 
\ee 
Then the action of $\lrb Q_0,Q_+,Q_- \rrb$ defined by (\ref{q2+bose}) on these basis 
states leads to the irreducible representation (\ref{q2+rep}).  

Now, we can make the association
\be
\lmd j,m,l \rra \lra \fr{z_2^{2j}z_3^{-(2l+j)}\lrb z_1z_3/z_2\rrb^{j+m}}
                     {\sqrt{(j+m)!(j-m)!(m-2l)!}}\,.
\ee
Since $j$ and $l$ are constants, this shows that the set of functions
\be
\psi^+_{j,l,n}(z) = \fr{z^n}{\sqrt{n!(2j-n)!(n-2l-j)!}}\,, \quad 
 n = 0,1,2,\,\ld\,,2j\,, 
\ee
forms the basis for the representation (\ref{q2+rep}) corresponding to the 
single variable realization 
\be
Q_0 = z\fr{d}{dz}-j-l\,, \quad
Q_+ = -z^2\fr{d}{dz} + 2jz\,, \quad
Q_- = z\fr{d^2}{dz^2} - (2l-j-1)\fr{d}{dz}\,.
\ee 

\section{Representations of $Q^-(1,1)$} 
\renewcommand{\theequation}{\arabic{section}.{\arabic{equation}}}
\setcounter{equation}{0}
For the algebra $Q^-(1,1)$ defined in (\ref{q11-}) the condition that $\ck$ 
and 
$\cl$ take constant values in an irreducible representation fixes the basis to 
be the set of states 
\be
\lmd k , l , n \rra = \lmd k,n \rra \lmd 2l-k-n \rra\,, \quad 
 n = 0, 1, 2,\,\ld\,, (2l-k)\,, 
\lb{q11-base}
\ee
where $\lcb\lmd k,n \rra\rcb$ are the $su(1,1)$ basis states (\ref{su11rep}), 
$\lcb\lmd 2l-k-n \rra\rcb$ are the oscillator states such that 
$N_3\lmd 2l-k-n \rra$ $=$ $(2l-k-n)\lmd 2l-k-n \rra$, 
\be
2l-k = 0,1,2,\,\ld\,, \qquad k = 1/2, 1, 3/2,\,\ld\,, 
\ee  
and 
\be 
\ck \lmd k , l , n \rra = k(1-k) \lmd k , l , n \rra\,, \quad 
\cl \lmd k , l , n \rra = l \lmd k , l , n \rra\,. 
\ee 
The basis states (\ref{q11-base}) carry the $(k,l)$-th, $(2l-k+1)$-dimensional,  
unitary irreducible representation\,:   
\bea 
Q_0 \lmd k , l , n \rra & = & (k-l+n) \lmd k , l , n \rra\,, \nn \\ 
Q_+ \lmd k , l , n \rra & = & 
  \sqrt{(n+2k)(n+1)(2l-k-n)}\,\lmd k , l , n+1 \rra\,, \nn \\
Q_- \lmd k , l , n \rra & = & 
  \sqrt{(n+2k-1)n(2l-k-n+1)}\,\lmd k , l , n-1 \rra\,. \nn \\ 
 &  & \qquad \qquad n = 0,1,2,\,\ld\,,(2l-k)\,. 
\lb{q11-rep} 
\eea 
The Casimir operator (\ref{q11-co}) has the value 
$(l+1)\lsb k(1-k) + l(l-1)\rsb$ in this representation.  

For this algebra there is a two dimensional representation for each value of 
$k = 1/2,1,3/2,\ld\,,$ as given by 
\bea 
Q_0 & = & \half \lrb \ba{cc}
                     k-1 & 0   \\
                     0   & k+1 \ea \rrb\,, \quad 
Q_+ = \lrb \ba{cc} 
                   0 & 0 \\
           \sqrt{2k} & 0 \ea \rrb\,, \quad 
Q_- = \lrb \ba{cc}
          0 & \sqrt{2k}  \\
          0 & 0          \ea \rrb\,, \nn \\ 
\ck & = & k(1-k)\,, \quad 
\cl = l = \frac{1}{2}(k+1)\,, \quad   
\cc = \frac{1}{8}\lrb -3k^3 - 5k^2 + 11k - 3 \rrb\,, \nn \\
  &   &  
\eea
as can be verified directly. 

By using the two-mode bosonic realization of $su(1,1)$ we can write down the 
three-mode bosonic realization of $Q^-(1,1)$ as 
\be
Q_0 = \fr{1}{4} \lrb a_1^\da a_1 + a_2^\da a_2 
           - 2a_3^\da a_3 + 1 \rrb\,, \quad  
Q_+ = a_1^\da a_2^\da a_3\,, \quad 
Q_- = Q_+^\da = a_1 a_2 a_3^\da\,.  
\lb{q11-bose}
\ee
Translating the basis states (\ref{q11-base}) into the three-mode Fock states it 
is found that the action of $\lrb Q_0, Q_+, Q_- \rrb$ defined by (\ref{q11-bose}) 
on the basis states 
$\lcb\lmd k,l,n \rra\right.$ $=$ 
$\left.\lmd n+2k-1 \rra \lmd n \rra \lmd 2l-k-n \rra\rcb$ 
leads to the representation (\ref{q11-rep}).  

As before, let us make the association 
\be
\lmd k , l , n \rra \lra \fr{z_1^{2k-1} z_3^{2l-k} \lrb z_1z_2/z_3 \rrb^n}
                            {\sqrt{(n+2k-1)!n!(2l-k-n)!}}\,. 
\ee
Since $k$ and $l$ are constants for a given representation we can take 
\be
\phi^-_{k,l,n}(z) = \fr{z^n}{\sqrt{(n+2k-1)!n!(2l-k-n)!}}\,, \quad 
  n = 0,1,2,\,\ld\,,(2l-k)\,, 
\ee
as the set of basis functions for the single variable realization 
\be 
Q_0 = z \ddz + k - l\,, \quad 
Q_+ = -z^2 \ddz + (2l-k)z\,, \quad 
Q_- = z \ddzt + 2k \ddz\,, 
\ee
leading to the representation (\ref{q11-rep}).  

\section{Representations of $Q^+(1,1)$} 
\renewcommand{\theequation}{\arabic{section}.{\arabic{equation}}}
\setcounter{equation}{0}
For the algebra $Q^+(1,1)$ defined in (\ref{q11+}) the constancy of $\ck$ and 
$\cl$ in any irreducible representation fixes the basis to be the set of states 
\be
\lmd k , l , n \rra = \lmd k,n \rra \lmd n+k-2l \rra\,, \quad 
 n = 0, 1, 2,\,\ld\,, 
\lb{q11+base}
\ee
where $\lcb\lmd k,n \rra\rcb$ are the $su(1,1)$ basis states, 
$\lcb\lmd n+k-2l \rra\rcb$ are the oscillator states such that 
$N_3\lmd n+k-2l \rra$ $=$ $(n+k-2l)\lmd n+k-2l \rra$, 
\be
k-2l = 0,1,2,\ld\,, \quad k = 1/2, 1, 3/2, \ld\,, 
\ee
and 
\be
\ck \lmd k,l,n \rra = k(1-k) \lmd k,l,n \rra\,, \qquad
\cl \lmd k,l,n \rra = l \lmd k,l,n \rra\,. 
\ee 
The set of basis states (\ref{q11+base}) carry the $(k,l)$-th, infinite dimensional 
unitary irreducible representation\,: 
\bea 
Q_0 \lmd k,l,n \rra & = & (k-l+n) \lmd k,l,n \rra\,, \nn \\ 
Q_+ \lmd k,l,n \rra & = & \sqrt{(n+2k)(n+1)(n+k-2l+1)}\,\lmd k,l,n+1 \rra\, \nn \\  
Q_- \lmd k,l,n \rra & = & \sqrt{(n+2k-1)n(n+k-2l)}\,\lmd k,l,n-1 \rra\,, \nn \\ 
   &   & \qquad \qquad n = 0,1,2,\ld\,. 
\lb{q11+rep}
\eea 
The Casimir operator (\ref{q11+co}) has the value $l\lrb l-k^2 \rrb$ in this 
representation. 

In terms of three bosonic modes the realization of $\lrb Q_0, Q_+, Q_- \rrb$ 
is given by 
\be
Q_0 = \fr{1}{4} \lrb a_1^\da a_1 + a_2^\da a_2 
           + 2a_3^\da a_3 + 1 \rrb\,, \quad  
Q_+ = a_1^\da a_2^\da a_3^\da\,, \quad 
Q_- = Q_+^\da = a_1 a_2 a_3\,. 
\lb{q11+bose} 
\ee 
Application of this realization on the set of three-mode Fock states 
\be
\lmd k , l , n \rra = \lmd n+2k-1\rra \lmd n \rra \lmd n+k-2l \rra\,, \quad 
 n = 0, 1, 2,\,\ld\,, 
\ee
leads to the $(k,l)$-th irreducible representation (\ref{q11+rep}).  

From the association 
\be 
\lmd k , l , n \rra \lra \fr{z_1^{2k-1} z_3^{k-2l}\lrb z_1 z_2 z_3 \rrb^n} 
                         {\sqrt{(n+2k-1)!n!(n+k-2l)!}} 
\ee 
it is clear that we can take 
\be 
\phi^+_{k,l,n}(z) = \fr{z^n}{\sqrt{(n+2k-1)!n!(n+k-2l)!}} 
\ee
as the set of basis functions for the single variable realization associated 
with the representation (\ref{q11+rep}).  The corresponding realization is 
\bea
Q_0 & = & z \ddz + k - l\,, \qquad Q_+ = z\,, \nn \\  
Q_- & = & z^2 \ddzth + (3k-2l+2) z \ddzt + \lrb 2k^2-4kl+2k \rrb \ddz\,. 
\eea 

\section{Concluding remarks} 
\renewcommand{\theequation}{\arabic{section}.{\arabic{equation}}}
\setcounter{equation}{0}

\noindent
As already mentioned in the introduction, polynomial algebras occur in 
several physical problems and have mathematically rich structures.  Here we 
shall like to make some observations with reference to the three dimensional 
quadratic algebras we have constructed.  

Let us first look at the Dicke model in quantum optics.  This model, 
generalizing the Jaynes-Cummings model, describes the interaction of the 
radiation field wtih a collection of identical two-level atoms located 
within a distance much smaller than the wavelength of the radiation.  In 
the particular case when the atoms interact resonantly with a single mode 
coherent cavity field, the (Tavis-Cummings) Hamiltonian, under the electric 
dipole and rotating wave approximations, is given by (in the units $\hbar = 1$) 
\be
H = \omega\lrb J_0+a^\da a \rrb + gJ_+a + g^*J_-a^\da\,. 
\lb{tch} 
\ee
where $\omega$ is the frequency of the field mode (and the atomic transition), 
$J_0+a^\da a$ is the excitation number operator, and $g$ is the coupling 
constant.  The annihilation and creation operators $a$ and $a^\da$ correspond 
to the single mode radiation field.  The operators 
\be
J_0 = \sum_j = \sigma^j_0\,, \qquad
J_\pm = \sum_j = \sigma^j_\pm\,,
\ee
with $\sigma^j_{0,\pm}$ as mutually commuting triplets of the Pauli matrices, 
obey the $su(2)$ algebra and define the collective atomic operators.  From 
purely physical arguments it is possible to construct the matrix representation 
of the Hamiltonian (\ref{tch}) and study its spectrum and applications  
\cite{TC,WB} (see \cite{CK} for an extensive review).  The excitation number 
operator (corresponding to our $\cl$) and $J^2$ are known integrals of motion 
for the system.  Thus it is now easy to recognize $Q^-(2)$ as the dynamical 
algebra of the Hamiltonian (\ref{tch}).  The two cases of representations 
(\ref{q2-rep1}-\ref{q2-rep2}) we have found are exactly the ones identified 
in the literature from a physical point of view.  We hope that this precise 
identification of the dynamical algebra of the Dicke model would help its further 
understanding.  In this regard, our earlier proposal of a general method for 
constructing the Barut-Girardello-type and Perelomov-type coherent states of any 
three dimensional polynomial algebra \cite{Su} should be useful. In \cite{Su} we 
have, in particular, considered the examples of $Q^{\pm}(1,1)$ and the cubic 
Higgs algebra with reference to the construction of coherent states.      

One should also note the following.  As in the case of any Lie algebra one may 
also present a polynomial algebra equivalently by choosing linear combinations of 
the generators as new generators.  For example, following \cite{DYSK}, we may 
take $X_3 = Q_0+\cl$, $X_+ = J_+a$ and $X_- = J_-a^\da$ as the three generators 
of $Q^-(2)$.  Then the corresponding algebraic relations are   
\be 
\lsb X_3 , X_\pm \rsb = \pm X_\pm\,, \qquad 
\lsb X_+ , X_- \rsb = -3X_3^2 + (4\cl+1)X_3 + J^2\,. 
\lb{newalg} 
\ee 
This is easily seen by substituting $Q_0 = X_3-\cl$ in (\ref{q2-}).  In 
\cite{DYSK} the three dimensional quadratic algebra defined by (\ref{newalg}), 
equivalent to $Q^-(2)$, has been considered for constructing the relative-phase 
operator for the Dicke model.  

Next let us consider the Hamiltonian 
\be
H = a_1^\da a_1 + a_2^\da a_2 + 2a_3^\da a_3 + 2 
\lb{3daniso}
\ee 
which describes, in the units $\hbar = 1$ and $\omega = 1$, a three dimensional 
anisotropic quantum harmonic oscillator with the frequency in the third direction 
twice that in the perpendicular plane.  If we relate $H$ to $Q_0$ in 
(\ref{q11+bose}) it is clear that $Q^+(1,1)$ is the dynamical algebra of the 
system (\ref{3daniso}).  From the representations (\ref{q11+rep}) one can easily 
arrive at the result that the spectrum of $H$ is the set of all integers 
$\geq 2$.  Let us look at the invariance algebra of $H$.  From 
the construction of $Q^-(1,1)$ we recognize that 
\be
H = 4\cl + 1\,, 
\ee
where $\cl$ is a central element of the algebra generated by 
$\lrb Q_0, Q_\pm \rrb$ in (\ref{q11-}) or (\ref{q11-bose}).  Thus, 
$\lrb \ck , Q_0 , Q_\pm \rrb$ are the integrals of motion for the system 
(\ref{3daniso}), or in other words, $Q^-(1,1)$ is the invariance algebra of the 
system.  Since $\cl$ has the spectrum 
\be
\cl = l = n/4\,, \qquad n = 1,2,3,\ld\,, 
\ee
it is clear that the Hamiltonian (\ref{3daniso}) has the spectrum 
\be
H = N+2\,, \qquad N = 0,1,2,\ld\,. 
\ee
Each level can be labeled by the eigenvalues of a complete set of commuting 
operators $\lrb H-2 , \ck \rrb$.  It is interesting to compute the 
degeneracy of the $N$-th level using the representation theory of the algebra 
(\ref{q11-}).  For the $N$-th level the value of $\cl$ is $l = (N+1)/4$.  
Calculating the corresponding values of $k$ for which finite dimensional 
representations are possible we find that the dimensions of the associated 
irreducible representations are $(1,2,\ld\,,2m+1)$ if $N = 4m$ or $4m+1$\,, 
and $(1,2,\ld\,,2m+2)$ if $N = 4m+2$ or $4m+3$.  The degeneracy of the level 
is the sum of the dimension of the $k = 1/2$ representation and twice the 
dimensions of $k > 1/2$ representations.  One has to count the dimensions of 
$k > 1/2$ representations twice in the sum since there are two possible choices 
for the bases leading to the same representation in these cases as already 
noted.  Now, the four cases, $N = 4m, 4m+1, 4m+2$ and $ 4m+3$, are to be 
considered separately.  The result is as follows\,: the degeneracies of the 
levels, $N = 4m, 4m+1, 4m+2$ and $ 4m+3$, resectively, are $(2m+1)^2$, 
$(2m+1)(2m+2)$, $4(m+1)^2$ and $2(m+1)(2m+3)$\,.  In other words, the number 
of compositions of the integer $N$ (partitions with ordering taken into account) 
in the prescribed pattern $n_1 + n_2 + 2n_3$, with the interchange of $n_1$ and 
$n_2$ taken into account, is $(2m+1)^2$, $(2m+1)(2m+2)$, $4(m+1)^2$ and 
$2(m+1)(2m+3)$, if $N = 4m, 4m+1, 4m+2$, and $4m+3$, resectively.  It is to be 
noted that in this example the sum of all the dimensions of the irreducible 
representations associated with the given $l = (N+1)/4$ gives the number of 
partitions of $N$ in the pattern $n_1 + n_2 + 2n_3$, disregarding the 
interchange of $n_1$ and $n_2$.  This leads to the result that the number of 
such partitions is $(m+1)(2m+1)$ for $N = 4m$ or $4m+1$ and $(m+1)(2m+3)$ for 
$N = 4m+2$ or $4m+3$.  Thus, it is interesting to observe this connection 
between a three dimensional quadratic algebra and the theory of partitions. 

It should be noted that if one can identify a given three dimensional 
quadratic algebra as belonging to one of the four classes we have considered 
then its representation theory can be worked out immediately, at least 
partially.  For example, observe that 
\be
Q_0 = a^\da a\,, \qquad 
Q_+ = \fr{1}{\sqrt{3}}\lrb a^\da \rrb^3\,, \qquad 
Q_- = \fr{1}{\sqrt{3}}\,a^3\,,
\ee
obey the algebra 
\be
\lsb Q_0 , Q_\pm \rsb = \pm Q_\pm\,, \qquad 
\lsb Q_+ , Q_- \rsb = -3Q_0^2 - 3Q_0 + 2\,.
\ee 
This algebra is uniquely identified with $Q^+(1,1)$ with $\cl$ $=$ $l = 1$ 
and $\ck$ $=$ $k(1-k) = -2$ (or $k = 2$).  Correspondingly the algebra is 
seen to have the infinite dimensional representation given by 
\bea 
Q_0 \lmd n \rra & = & (n+1) \lmd n \rra\,, \nn \\
Q_+ \lmd n \rra & = & (n+1)\sqrt{n+4} \lmd n+1 \rra\,, \quad 
Q_- \lmd n \rra = n\sqrt{n+3} \lmd n-1 \rra\,, \nn \\
  &   &  \qquad \qquad \qquad \qquad \qquad n = 0,1,2,\,\ld\,\,. 
\eea
The fact that the representations we have discussed are not complete is clear 
from the following example. For a two dimensional anisotropic quantum harmonic 
oscillator the Hamiltonian is  
\be
H = a_1^\da a_1 + 2a_2^\da a_2 + \fr{3}{2}\,, 
\ee
in the units $\hbar = 1$ and $\omega = 1$.  The invariance algebra of this 
Hamiltonian is generated by 
\be
Q_0 = \fr{1}{4}\lrb a_1^\da a_1 - 2a_2^\da a_2 + \half \rrb\,,\quad 
Q_+ = \half\lrb a_1^\da \rrb^2 a_2\,, \quad 
Q_- = \half a_1^2 a_2^\da\,, 
\ee
and the algebra is given by 
\bea 
\lsb Q_0 , Q_\pm \rsb & = & \pm Q_\pm\,, \nn \\
\lsb Q_+ , Q_- \rsb & = & 3Q_0^2 + \half (H-3)Q_0 
                           - \fr{1}{16} H(H+2) + \fr{3}{8}\,. 
\lb{2daniso} 
\eea  
This algebra can be readily identified with $Q^-(1,1)$ corresponding to $\cl$ 
$=$ $(H-1)/4$ and $\ck = k(1-k) = 3/16$ or $k = 1/4$ or $3/4$.  It may be noted 
that the corresponding representations cannot be presented in terms of the 
three-boson Fock states since these, considered in Section 6, correspond only to 
$k$ $=$ $1/2,1,3/2,\,\ld\,$.  To get the representations of the algebra 
(\ref{2daniso}) one will have to combine the representations of a boson algebra 
with the representations of $su(1,1)$ for $k$ $=$ $1/4$ or $3/4$ (in terms of 
single boson Fock states). A detailed discussion of the algebraic approach to 
the two dimensional quantum system of an anisotropic oscillator with an 
additional singular potential in one direction is found in \cite{Le}.  

An interesting possibility is suggested by the structure of the algebra 
$Q^-(1,1)$.  Let us define 
\be
N = Q_0\,, \quad
A = \frac{1}{\sqrt{\cl(\cl+1)-\ck}}\,Q_-\,, \quad
A^\da = \frac{1}{\sqrt{\cl(\cl+1)-\ck}}\,Q_+\,.
\ee
Then the algebra (\ref{q11-}) becomes 
\bea
\lsb N , A \rsb & = & -A\,, \quad 
\lsb N , A^\da \rsb = A^\da\,, \nn \\ 
\lsb A , A^\da \rsb & = & 1 - \fr{2\cl-1}{\cl(\cl+1)-\ck}\,N
                            - \fr{3}{\cl(\cl+1)-\ck}\,N^2\,.  
\lb{qo}
\eea 
We may consider this as the defining algebra of a quadratic oscillator, 
corresponding to a special case of the general class of deformed oscillators 
(\cite{Ar}-\cite{D})\,:
\be
\lsb N , A \rsb = -A\,, \quad 
\lsb N , A^\da \rsb = A^\da\,, \quad
\lsb A , A^\da \rsb = F(N)\,.
\ee
The quadratic oscillator (\ref{qo}) belongs to the class of generalized 
deformed parafermions \cite{Q}.  It should be interesting to study the 
physics of assemblies of quadratic oscillators.  In fact, the canonical 
fermion, with 
\be 
N = \lrb \ba{cc}
         0 & 0  \\
         0 & 1 \ea \rrb\,, \quad 
f = \lrb \ba{cc} 
         0 & 1 \\
         0 & 0 \ea \rrb\,, \quad 
f^\da = \lrb \ba{cc}
          0 & 0  \\
          1 & 0  \ea \rrb\,, 
\ee
is a quadratic oscillator!  Observe that 
\be
\lsb N , f \rsb = -f\,, \quad 
\lsb N , f^\da \rsb = f^\da\,, \quad
\lsb f , f^\da \rsb = 1 - \half N - \fr{3}{2} N^2\,.
\ee

To conclude, starting with $su(2)$, $su(1,1)$, and an oscillator algebra, we 
have constructed four classes of three dimensional quadratic algebras of 
the type 
\be
\lsb Q_0 , Q_\pm \rsb = \pm Q_\pm\,, \quad 
\lsb Q_+ , Q_- \rsb = a Q_0^2 + b Q_0 + c\,.  
\ee  
In each class the structure constants $(a,b,c)$ take particular series of 
values.  We have also found for these algebras the three-mode bosonic 
realizations, corresponding matrix representations, and single variable 
differential operator realizations.  We hope that this work would lead to 
further insights into a more complete understanding of the representation 
theory of arbitrary three dimensional quadratic algebras.  \\ 

\section*{Acknowledgements}
One of us (VSK) is thankful to the Institute of Mathematical Sciences, Chennai, 
for hospitality and financial support for the period of his stay there during 
which this work was completed.  We would like to thank the referees for very 
constructive comments and suggestions which improved the presentation of this 
paper.  \\

\end{document}